\documentclass[preprint,prd,nofootinbib,tightenlines,amsmath]{revtex4}

\usepackage{epsfig}
\usepackage{dcolumn}
\usepackage{bm}

\begin{document}

\baselineskip=15pt

\preprint{hep-ph/0509041}

\preprint{WSU-HEP-0505}

\hspace*{\fill} $\hphantom{-}$

\title{Implications of a New Particle from \\
the HyperCP Data on $\bm{\Sigma^{+}\to p \mu^{+}\mu^{-}}$}

\author{Xiao-Gang He}

\email{hexg@phys.ntu.edu.tw}

\affiliation{Department of Physics, Nankai University, Tianjin\\}

\affiliation{NCTS/TPE, Department of Physics,\\ National Taiwan
University, Taipei}

\author{Jusak Tandean}

\email{jtandean@ulv.edu}

\affiliation{Department of Physics and Astronomy, Wayne State University,
Detroit, MI 48201\\}

\affiliation{Department of Mathematics/Physics/Computer Science,
University of La Verne, La Verne, CA 91750\footnote{Present address}}

\author{G. Valencia}

\email{valencia@iastate.edu}

\affiliation{Department of Physics and Astronomy, Iowa State University, Ames, IA 50011\\}

\date{\today}

\begin{abstract}
The HyperCP collaboration has recently reported the observation of
three events for the decay  $\Sigma^{+}\to p \mu^{+}\mu^{-}$ with
an invariant mass $m_{\mu^+\mu^-}$ for the muon-antimuon pair of
\,$\sim$214\,MeV.  
They suggest that a new particle state $X$
may be needed to explain the observed $m_{\mu^+\mu^-}$ distribution.
Motivated by this result, we study the properties of such a
hypothetical particle. We first use $K^+\to \pi^+\mu^+\mu^-$ data to
conclude that $X$ cannot be a scalar or vector particle.
We then collect existing constraints on a pseudoscalar or
axial-vector $X$ and find that these possibilities are still allowed
as explanations for the HyperCP data. Finally we assume that the
HyperCP data is indeed explained by a new pseudoscalar or
axial-vector particle and use this to predict enhanced rates for
$K_L\to \pi\pi X\to\pi\pi\mu^+\mu^-$  and
$\Omega^-\to\Xi^- X\to\Xi^-\mu^+\mu^-$.
\end{abstract}

\pacs{PACS numbers: }

\maketitle

\section{Introduction}

Three events for the decay mode \,$\Sigma^{+}\to p \mu^{+}\mu^{-}$\,
with an invariant mass of \,$214.3\pm 0.5$\,MeV\, for the
muon-antimuon pair have been recently observed by the HyperCP
collaboration~\cite{Park:2005ek}. The branching ratio is obtained to
be $\bigl[8.6^{+6.6}_{-5.4}(\rm stat)\pm 5.5(syst)\bigr]\times
10^{-8}$~\cite{Park:2005ek}. The central value is considerably
larger than the short-distance contribution in the standard
model~\cite{He:2005yn}. When long-distance contributions are
properly included, it is possible to account for the total branching
ratio~\cite{He:2005yn,Bergstrom:1987wr}. However, the clustering of
the events $\bigl[$whose contribution to the branching ratio is
$\bigl(3.1^{+2.4}_{-1.9}\pm 1.5\bigr)\times 10^{-8}\bigr]$ around
214\,MeV cannot be explained. If this result stands future
experimental scrutiny, it is most likely to be due to a particle
state $X$ having a mass of 214\,MeV. In this paper we study the
properties of such a particle assuming its existence.

The mass 214 MeV of this hypothetical particle is close to, but
higher than, the sum of the masses of two muons. It is tempting to
identify it as a muonium bound-state. However, the S-wave
bound-state has a mass below the sum of the two muon masses.
Therefore, the state $X$ cannot be an S-wave muonium state. Radial
excitations can yield larger masses, but it is unlikely that the
electromagnetic interaction bounding the muon and antimuon together
can raise the mass by the 3\,MeV needed. The $X$ particle, if
exists, is likely a new state beyond the standard model (SM). There
are theories where such light states naturally exist, for example,
the super-partner of the goldstino particle in spontaneously local
super-symmetry breaking theories as discussed in
Ref.~\cite{Gorbunov:2000cz}. These particles, pseudoscalar and
scalar ones, can have masses lower than a few GeV or even in the MeV
range.

In our study we will not attempt to construct models which predict
such particles. Instead, we will assume the existence of the new
particles and study the implications from the HyperCP data. To be
consistent with observations, we follow HyperCP and assume that the
hypothetical particles have small widths, are short-lived (they
decay inside the detector), and do not interact
strongly~\cite{Park:2005ek}.

\section{Effective interactions\label{effint}}

In our study, we will try to be as model independent as possible by
parameterizing the interactions of this new particle with known particles.
In the HyperCP hypothesis, this particle is produced in  the decay of
$\Sigma$ to $p$ and subsequently decays into a muon-antimuon pair.
At the quark level, the particle $X$ must then couple to
$\bar d s$  (and of course to $\mu^+\mu^-$ as well). A priori, the
state  $X$ can be a scalar, pseudoscalar, vector, axial-vector, or
even a tensor particle. We will consider four possibilities:
scalar ($X_S$), pseudoscalar ($X_P$), vector ($X_V$), and axial
vector ($X_A$).

Assuming that the hypothetical new particles have definite parity, do not carry
electric or color charge, and are their own anti-particles, we can write their
couplings to  $\bar d s$ and $\mu^+\mu^-$ as
\begin{eqnarray}
{\cal L}_S &=& \bigl(-g_{Sq}^{}  \bar d s \,+\,  {\rm H.c.} \bigr)X_S
\,+\,  g_{S\mu}^{}\bar \mu \mu X_S  \,\,,\nonumber\\
{\cal L}_P &=& \bigl(-ig_{Pq}^{} \bar d\gamma_5 s  \,+\,  {\rm H.c.} \bigr)X_P
\,+\,
ig_{P\mu}^{}\bar \mu \gamma_5 \mu X_P\,\,,\nonumber\\
{\cal L}_V &=& \bigl(-g_{Vq}^{} \bar d \gamma_\mu s  \,+\,  {\rm H.c.} \bigr) X^\mu_V
\,+\, g_{V\mu}^{}\bar \mu \gamma_\mu \mu X^\mu_V\,\,,\nonumber\\
{\cal L}_A &=& \bigl(g_{Aq}^{} \bar d \gamma_\mu \gamma_5 s \,+\, {\rm H.c.}\bigr)X^\mu_A
\,+\,  g_{A\mu}^{} \bar \mu \gamma_\mu \gamma_5 \mu X_A^\mu  \,\,.
\label{quark}
\end{eqnarray}
If the particle does not have a definite parity, our results should be
interpreted as applying to the parity-even or -odd coupling as appropriate.

A condition that the couplings in the above equations must satisfy
is that they must be able to produce the observed branching ratio
in  \,$\Sigma^+\to p\mu^+\mu^-$.\,  To carry out such a fit, one
must know how the $X$ couples to the hadron states $\Sigma^+$ and
$p$ from the above quark-level couplings. To this end, we employ
chiral perturbation theory to obtain the couplings. Our task is simplified
by the assumption that the hypothetical particles do not interact strongly
as they can then be readily identified with the scalar, pseudoscalar, vector,
and axial-vector external sources in the standard-model Lagrangians. With
the flavor properties assumed in Eq.~(\ref{quark}), the appropriate
Lagrangians are then
\begin{subequations}   \label{eff}
\begin{eqnarray}   \label{L_SH}
{\cal L}_{SB\varphi}^{}  &=& b_D^{} \left\langle \bar{B}{}^{}
\left\{ \xi^\dagger h_S^{}\xi^\dagger+\xi h_S^{}\xi, B^{} \right\}
\right\rangle + b_F^{} \left\langle \bar{B}{}^{}
 \left[ \xi^\dagger h_S^{}\xi^\dagger+\xi h_S^{}\xi, B^{} \right]
\right\rangle \nonumber \\ && +\,\, b_0^{} \left\langle h_S^{}\,
\bigl(\Sigma^\dagger+\Sigma\bigr) \right\rangle \left\langle
\bar{B}{}^{} B^{} \right\rangle \,+\, \frac{1}{2} f^2 B_0^{}
\left\langle h_S^{}\, \bigl(\Sigma^\dagger+\Sigma\bigr)
\right\rangle \,\,+\,\,  {\rm H.c.}    \,\,,
\end{eqnarray}
\begin{eqnarray}   \label{L_PH}
{\cal L}_{PB\varphi}^{}  &=&  i  b_D^{} \left\langle
\bar{B}{}^{} \left\{ \xi^\dagger h_P^{}\xi^\dagger-\xi h_P^{}\xi,
B^{} \right\} \right\rangle +  i  b_F^{} \left\langle
\bar{B}{}^{}
 \left[ \xi^\dagger h_P^{}\xi^\dagger-\xi h_P^{}\xi, B^{} \right]
\right\rangle \nonumber \\ && +\,\,  i  b_0^{} \left\langle h_P^{}\,
\bigl(\Sigma^\dagger-\Sigma\bigr) \right\rangle \left\langle
\bar{B}{}^{} B^{} \right\rangle \,\,+\,\, \frac{ i }{2} f^2 B_0^{}
\left\langle h_P^{}\, \bigl(\Sigma^\dagger-\Sigma\bigr)
\right\rangle \,\,+\,\,  {\rm H.c.}    \,\,,
\end{eqnarray}
\begin{eqnarray}   \label{L_VH}
{\cal L}_{VB\varphi}^{}  &=& \frac{1}{2} \Bigl\langle \bar{B}
\gamma_\mu^{}\, \bigl[ B,\, \xi^\dagger h_V^\mu\xi+\xi
h_V^\mu\xi^\dagger \bigr] \Bigr\rangle \nonumber \\ && +\,\,
\frac{1}{2} D\, \Bigl\langle \bar{B}\gamma_\mu^{}\gamma_5^{}\,
\bigl\{ \xi^\dagger h_V^\mu\xi-\xi h_V^\mu\xi^\dagger,\, B \bigr\}
\Bigr\rangle \,+\, \frac{1}{2} F\, \Bigl\langle
\bar{B}\gamma_\mu^{}\gamma_5^{}\, \bigl[ \xi^\dagger
h_V^\mu\xi-\xi h_V^\mu\xi^\dagger,\, B \bigr] \Bigr\rangle
\nonumber \\ && +\,\, \frac{1}{2}\, {\cal C} \left[\,
\bar{T}_\mu^{} \left( \xi^\dagger h_V^\mu\xi-\xi
h_V^\mu\xi^\dagger \right) B + \bar{B} \left( \xi^\dagger
h_V^\mu\xi-\xi h_V^\mu\xi^\dagger \right) T_\mu^{} \,\right]
\nonumber \\ && -\,\, \frac i {2} f_{}^2\, \bigl\langle
h_V^\mu\, \bigl(
\partial_\mu^{}\Sigma\, \Sigma^\dagger-\Sigma^\dagger\, \partial_\mu^{}\Sigma
\bigr) \bigr\rangle \,\,+\,\,  {\rm H.c.}   \,\,,
\end{eqnarray}
\begin{eqnarray}   \label{L_AH}
{\cal L}_{AB\varphi}^{}  &=& \frac{ 1}{2} \Bigl\langle
\bar{B}\, \gamma_\mu^{}\, \Bigl[ B,\, \xi^\dagger h_A^\mu\xi-\xi
h_A^\mu\xi^\dagger \Bigr] \Bigr\rangle \nonumber \\ && +\,\,
\frac{1}{2} D\, \Bigl\langle \bar{B}\gamma_\mu^{}\gamma_5^{}\,
\Bigl\{ \xi^\dagger h_A^\mu\xi+\xi h_A^\mu\xi^\dagger,\, B \Bigr\}
\Bigr\rangle \,+\, \frac{1}{2} F\, \Bigl\langle
\bar{B}\gamma_\mu^{}\gamma_5^{}\, \Bigl[ \xi^\dagger
h_A^\mu\xi+\xi h_A^\mu\xi^\dagger,\, B \Bigr] \Bigr\rangle
\nonumber \\ && +\,\, \frac{1}{2}_{}^{}\, {\cal C} \left[\,
\bar{T}_\mu^{} \left( \xi^\dagger h_A^\mu\xi+\xi
h_A^\mu\xi^\dagger \right) B + \bar{B} \left( \xi^\dagger
h_A^\mu\xi+\xi h_A^\mu\xi^\dagger \right) T_\mu^{} \,\right]
\nonumber \\ && -\,\, \frac i {2} f_{}^2\, \bigl\langle  h_A^\mu\,
\bigl( \partial_\mu^{}\Sigma\, \Sigma^\dagger+\Sigma^\dagger\,
\partial_\mu^{}\Sigma \bigr)
\bigr\rangle \,\,+\,\,  {\rm H.c.}   \,\,,
\end{eqnarray}
\end{subequations}
where we have shown only the terms relevant for this paper, and used the notation
$\,\bigl(h_Y^{}\bigr){}_{kl}^{}=X_{Y}\, g_{Yq}^{}\, h_{kl}^{}\,$  for
$\,Y=S,P,\,$  and $\,\bigl(h_Y^\mu\bigr){}_{kl}^{}=X_{Y}^\mu\,
g_{Yq}^{}\, h_{kl}^{}\,$  for  $\,Y=V,A,\,$ with
$\,h_{kl}^{}=(T_6^{}+{ i }T_7^{})_{kl}^{}=\delta_{k2}^{}\delta_{3l}^{}.\,$ The notation
and parameter values that we employ here are explained in
Appendix~\ref{Leff}. With the above effective Lagrangians, we can
obtain constraints on the couplings $g_{Yq}^{}$ from other low-energy
processes.

\section{Ruling out the scalar and vector as candidate particles\label{ruling}}

With the assumption that the new particles are short-lived and narrow,
their contribution to the branching ratio of  \,$\Sigma^+\to p\mu^+\mu^-$\,
is given by  \,${\cal B}(\Sigma \to p X){\cal B}(X\to \mu^+\mu^-)$.\,
Using the effective Lagrangians in Eq. (\ref{eff}), we find the matrix
elements for  \,$\Sigma^+\to p X$\,  to be
\begin{eqnarray}
{\cal M}(\Sigma^+\to p X_S) &=& -2g_{Sq}^{}\,(b_D-b_F)\, \bar p \Sigma^+ \,\,,
\nonumber\\
{\cal M}(\Sigma^+\to p X_P) &=& g_{Pq}^{}B_0\, (D-F)\,
\frac{m_\Sigma + m_p}{m^2_K - m^2_P}\, \bar p \gamma_5 \Sigma^+  \,\,,
\nonumber\\
{\cal M}(\Sigma^+\to p X_V) &=& -g_{Vq}^{}\, \bar p \gamma^{\mu}\Sigma^+ \epsilon^*_\mu  \,\,,
\nonumber\\
{\cal M}(\Sigma^+\to p X_A) &=& -g_{Aq}^{}\,(D-F)\, \bar p\gamma^{\mu}
\gamma_5 \Sigma^+\epsilon^*_\mu  \,\,.
\label{matrix}
\end{eqnarray}
These expressions follow from a kaon-pole diagram for the pseudoscalar,
and from a direct vertex from Eq.~(\ref{eff}) for the rest.
For the branching ratios, it then follows that
\begin{eqnarray}
{\cal B}(\Sigma^+\to p X_S\to p \mu^+\mu^-) &=&
9.0 \times10^{12}\, |g_{Sq}^{}|^2\, {\cal B}(X_S\to \mu^+\mu^-)  \,\,,
\nonumber\\
{\cal B}(\Sigma^+\to p X_P\to p \mu^+\mu^-) &=&
3.7 \times 10^{11}\, |g_{Pq}^{}|^2\, {\cal B}(X_P\to \mu^+\mu^-) \,\,,
\nonumber\\
{\cal B}(\Sigma^+\to p X_V\to p \mu^+\mu^-) &=&
7.0 \times10^{11}\, |g_{Vq}^{}|^2\, {\cal B}(X_V\to \mu^+\mu^-) \,\,,
\nonumber\\
{\cal B}(\Sigma^+\to p X_A\to p \mu^+\mu^-) &=&
7.0 \times10^{11}\, |g_{Aq}^{}|^2\, {\cal B}(X_A\to \mu^+\mu^-)  \,\,.
\label{sigmar}
\end{eqnarray}

For the scalar and vector particles, there are severe constraints
from \,$K^{\pm}\to \pi^{\pm} \mu^+\mu^-$.\, The branching ratio of
\,$K^{\pm}\to \pi^{\pm}\mu^+ \mu^-$\,  has been measured to be
\,${\cal B} = (8.1\pm1.4)\times 10^{-8}$~\cite{pdg}. The
$X$-particle contribution to these decays can again be factorized as
\,${\cal B}(K^{\pm}\to \pi^{\pm} X) {\cal B}(X\to \mu^+\mu^-)$.\,
Using the effective Lagrangians in Eq. (\ref{eff}), we have the
matrix elements for  \,$K^{\pm}\to \pi^{\pm} X$\,
\begin{eqnarray}
{\cal M}(K^{\pm} \to \pi^{\pm} X_S) &=& g_{Sq}^{} B_0  \,\,,\nonumber\\
{\cal M}(K^{\pm} \to \pi^{\pm} X_V) &=& g_{Vq}^{} (p_K +p_\pi)\cdot\epsilon^*  \,\,.
\end{eqnarray}
We have assumed $CP$ conservation for simplicity, and so taken the couplings
$g_{(S,V)q}^{}$ to be real.

The decay modes  \,$K^{\pm}\to\pi^{\pm}\mu^+\mu^-$\,  are long-distance
dominated in the SM~\cite{D'Ambrosio:1998yj} and the measured
spectra agree reasonably well with the
predictions~\cite{Ma:1999uj,Park:2001cv}. In particular, there is no
apparent bump in the  \,$m_{\mu^+\mu^-}=214$\,MeV\,
region~\cite{park-talk}. In view of this, we require that any
contribution from the hypothetical new particles to these rates be
below the experimental error, that is~\cite{pdg}
\begin{equation}
{\cal B}(K^{\pm}\to \pi^{\pm} \mu^+\mu^-)_{X}  \,\,\leq\,\, 1.4 \times 10^{-8}  \,\,.
\label{kpimmb}
\end{equation}
This leads to the constraints
\begin{eqnarray}
|g_{Sq}^{}|^2\, {\cal B}(X_S\to \mu^+\mu^{-}) \,\,<\,\,6.5\times 10^{-24}
\,\,,  \hspace{3em}
|g_{Vq}^{}|^2\, {\cal B}(X_V\to \mu^+\mu^-) \,\,<\,\, 4.3\times 10^{-23}  \,\,.
\end{eqnarray}
Combining these limits with Eq.~(\ref{sigmar}), we find
\begin{eqnarray}
{\cal B}(\Sigma^+\to p X_S\to p \mu^+\mu^-)  \,\,<\,\,  6\times10^{-11}  \,\,,
\hspace{3em}
{\cal B}(\Sigma^+\to p X_V\to p \mu^+\mu^-) \,\,<\,\,  3\times 10^{-11}  \,\,.
\end{eqnarray}
These results indicate that \,$K^{\pm}\to \pi^{\pm} \mu^+\mu^-$\, data rule
out both a scalar particle and a vector particle as explanations for the
HyperCP result. Notice that this conclusion still holds if we relax
Eq.~(\ref{kpimmb}) and allow the new contribution to be as large as the full
experimental rate.

The decays  \,$K^\pm\to\pi^\pm X_{P,A}$\,  are not allowed, as we have assumed
$X_{P,A}^{}$ to have no parity-odd couplings. Therefore, there
are no constraints from  \,$K\to\pi\mu^+\mu^-$\,  on the pseudoscalar and
axial-vector couplings of the hypothetical particles to quarks.

\section{Some constraints on pseudoscalar and axial-vector couplings\label{constr}}

We now consider other possible constraints on the  couplings
involving the pseudoscalar and axial-vector particles. We begin
by ignoring $CP$ violation so that $g_{Pq}^{}$ and $g_{Aq}^{}$ are real.
A strong constraint on flavor-changing neutral currents (FCNC) of this
type comes from  $K^{0}$-$\bar{K}^{0}$ mixing.
The mixing parameter  \,$M_{12}={\cal M}(K^0\to X\to\bar K^0)/2m_K$\,
from an intermediate $X$-state in  \,$K^0\to X\to\bar K^0$\, is given by
\begin{eqnarray}   \label{MKK}
{\cal M}(K^0\to X_P\to\bar K^0) &=&
\frac{2B^2_0 f^2\,g_{Pq}^2}{m^2_K- m^2_P}  \,\,,\nonumber\\
{\cal M}(K^0\to X_A\to \bar K^0) &=&  \frac{2 f^2\, g_{Aq}^2\, m^2_K}{m_A^2}  \,\,.
\end{eqnarray}
The measured value of \,$\Delta
M_{K_{L}-K_{S}}=3.483\times10^{-12}$\,MeV~\cite{pdg} can be
accommodated in the SM, but its calculation suffers from hadronic
uncertainties due to long-distance contributions. To be
conservative, we will thus require that any new physics contribution
be smaller than the experimental value, namely
\begin{equation}
\bigl(\Delta M_{K_{L}-K_{S}}\bigr)_{X}  \,\,=\,\,
2\ ({\rm Re}\,M_{12})_{X}  \,\,<\,\, 3.483 \times 10^{-12}\,\,\rm MeV  \,\,.
\end{equation}
With matrix elements from Eqs.~(\ref{L_PH}) and~(\ref{L_AH}), but
using  \,$f_{K}\sim 1.23f$,\, instead of $f$, in Eq.~(\ref{MKK})
for the kaon decay constant, this results in
\begin{eqnarray}
g_{Pq}^2  &<& 3.3\times 10^{-15}  \,\,,\nonumber\\
g_{Aq}^2  &<& 1.3\times 10^{-14}  \,\,.
\label{kbr}
\end{eqnarray}
When we substitute these bounds into Eq.~(\ref{sigmar}), we find
\begin{eqnarray}
\frac{{\cal B}(\Sigma^+\to p X_P\to p\mu^+\mu^-)}{{\cal B}(X_P\to\mu^+\mu^-)}
&<&  1.2\times10^{-3}  \,\,,\nonumber\\
\frac{{\cal B}(\Sigma^+\to p X_A\to p\mu^+\mu^-)}{{\cal B}(X_A\to\mu^+\mu^-)}
&<&  9.1\times10^{-3}  \,\,.
\end{eqnarray}
These constraints are so weak that  $X_{P}$ and $X_{A}$ are allowed candidates to
explain the HyperCP result, provided their branching ratios into muon pairs are
at least
\,${\cal B}(X_P\to \mu^+\mu^-) \geq 2.5 \times 10^{-5}$\,  and
\,${\cal B}(X_A\to \mu^+\mu^-) \geq 3.4 \times 10^{-6}$,\,  respectively.

If we allow for $CP$ violation in the $g_{Pq}^{}$ and $g_{Aq}^{}$ couplings,
the constraints from  $\Delta M_{K_{L}-K_{S}}$ become
\begin{eqnarray}
\bigl|({\rm Re}\,g_{Pq})^2-({\rm Im}\,g_{Pq})^2\bigr|
&<& 3.3 \times 10^{-15}  \,\,,\nonumber\\
\bigl|({\rm Re}\,g_{Pq})({\rm Im}\,g_{Pq})\bigr|
&<&  3.2\times 10^{-18}  \,\,,\nonumber\\
\bigl|({\rm Re}\,g_{Aq})^2-({\rm Im}\,g_{Aq})^2\bigr| &<& 1.3\times 10^{-14}
\,\,, \nonumber\\
\bigl|({\rm Re}\,g_{Aq})({\rm Im}\,g_{Aq})\bigr| &<& 1.2\times 10^{-17}  \,\,.
\label{kb}
\end{eqnarray}
The two additional constraints arise from the new-particle contribution to the
parameter  \,$\epsilon_K={\rm Im}\,M_{12}/\bigl(\sqrt{2}\,\Delta M_{K_{L}-K_{S}}\bigr)$.\,
This parameter can be calculated more reliably than  $\Delta M_{K_{L}-K_{S}}$ in
the standard model and is in good agreement with the result
\,$|\epsilon_K|=2.284\times 10^{-3}$.\,  In view of this,
we required the new-physics contribution to be less than  $30\%$ of the experimental
value, which is about the size of the theoretical uncertainty in the SM calculation.

New pseudoscalar and axial-vector particles also contribute to the rare decay
\,$K_L \to \mu^+\mu^{-}$\,  via the pole 
diagram  \,$K_L\to X\to\mu^+ \mu^-$.\, From Eqs.(\ref{quark})  and~(\ref{eff}), we obtain
\begin{eqnarray}
{\cal M}(K_L\to X_P\to\mu^+\mu^-) &=&  \frac{-2i B_0 f\,g_{Pq}^{}}{m^2_K-m^2_P}\,
g_{P\mu}^{}\bar\mu\gamma_5 \mu  \,\,,\nonumber \\
{\cal M}(K_L\to X_A\to\mu^+\mu^-) &=&
\frac{4 i f\, g_{Aq}^{}\, m_\mu^{}}{m^2_A}\, g_{A\mu}^{}\bar\mu\gamma_5\mu \,\,.
\end{eqnarray}
These matrix elements imply
\begin{eqnarray}
{\cal B}(K_L \to X_P\to \mu^+\mu^-) &=& 5.6\times 10^{18}\,\,{\rm GeV}^{-1}\,\,
g_{Pq}^2\,\, \Gamma(X_P\to\mu^+\mu^-)  \,\,,\nonumber\\
{\cal B}(K_L \to X_A\to \mu^+\mu^-) &=& 1.2\times 10^{18}\,\,{\rm GeV}^{-1}\,\,
g_{Aq}^2\,\, \Gamma(X_A\to\mu^+\mu^-)  \,\,.
\label{kltomm}
\end{eqnarray}
If we allow for $CP$ violation it is possible to obtain additional, weaker,
constraints from considering the mode  \,$K_{S}\to \mu^{+}\mu^{-}$.\,
To be useful, the equations above must be combined with additional information
on the couplings of the hypothetical new particles to muons. Partial information
can be obtained from considering their contribution to the anomalous magnetic
moment of the muon, $a_{\mu}$.

At one-loop level, the contributions of the new pseudoscalar and axial-vector to
$a_{\mu}$ are given respectively by
\begin{eqnarray}
a_\mu(P) &=&  -\frac{|g_{P\mu}^{}|^{2} m^2_\mu}{8\pi^2 m^2_P}\,
f_P\bigl(m^2_\mu/m^2_P\bigr)
\,\,=\,\, -2.28\times 10^{-3}\, |g_{P\mu}^{}|^2  \,\,,\nonumber\\
a_\mu(A) &=& \frac{|g_{A\mu}^{}|^{2} m^2_\mu}{4\pi^2 m^2_A}\,
f_A\bigl(m^2_\mu/m^2_A\bigr)
\,\,=\,\,  -8.97 \times 10^{-3}\, |g_{A\mu}^{}|^2   \,\,.
\end{eqnarray}
Here
\begin{eqnarray}
f_P(r) &=& \int^1_0 dx\, {x^3\over 1-x+rx^2}  \,\,,\nonumber\\
f_A(r) &=& \int^1_0 dx\, {4(x-1)x + x^2(1-x)-2r x^3\over 1-x +r x^2}  \,\,.
\end{eqnarray}

At present there is a discrepancy of $2.4\sigma$ between the SM
prediction and data~\cite{passera}, \,$\Delta
a_\mu=a_\mu(exp)-a_\mu(SM)=(23.9\pm 10)\times 10^{-10}$\, with
\,$a_\mu(exp)=(11659208\pm6)\times 10^{-10}$.\, We note that the new
contributions reduce the value of $a_{\mu}$, making the comparison
with experiment worse. In view of this, we place a conservative
constraint on $g^2_{X\mu}$ by requiring that the new contribution to
$a_{\mu}$ not exceed the experimental error. This results in
\begin{eqnarray}
|g_{P\mu}^{}|^2  \,\,<\,\, 2.6\times 10^{-7}  \,\,, & ~~ &
\Gamma(X_P \to \mu^+\mu^-) \,\,<\,\, 3.7\times 10^{-10}\,\,{\rm GeV}  \,\,,
\nonumber\\
|g_{A\mu}^{}|^2  \,\,<\,\, 6.7\times 10^{-8}  \,\,, & ~~ &
\Gamma(X_A \to \mu^+\mu^-) \,\,<\,\, 5.2\times 10^{-12}\,\, \rm GeV  \,\,.
\label{muonamm}
\end{eqnarray}
Combining the constraints in Eqs.~(\ref{kbr}) and~(\ref{muonamm}), we obtain
from Eq.~\ref{kltomm}
\begin{eqnarray}
{\cal B}(K_L \to X_P\to \mu^+\mu^-) &<& 6.8 \times 10^{-6}  \,\,,
\nonumber\\
{\cal B}(K_L \to X_A\to \mu^+\mu^-) &<&  8.1 \times 10^{-8}   \,\,.
\end{eqnarray}

The measured branching ratio for this mode,
${\cal B}(K_{L}\to \mu^{+}\mu^{-})= (6.87\pm 0.12)\times 10^{-9}$,
is almost completely saturated by the two-photon intermediate state,
the absorptive part of this contribution being
\,${\cal B}(K_{L}\to\gamma\gamma\to \mu^{+}\mu^{-})_{\rm abs}=
(6.63\pm 0.07)\times 10^{-9}$\,  (this is referred to as the unitarity bound).
This leaves little room for a direct new-physics contribution.
Here we assume that a possible new-physics contribution is at most equal to
the difference between the measured rate and the unitarity bound plus one
standard deviation,
\begin{equation}
{\cal B}(K_{L}\to \mu^{+}\mu^{-})_{X}  \,\,\leq\,\, 3.6 \times 10^{-10}  \,\,.
\end{equation}
Using this as a constraint improves the bounds of Eqs.~(\ref{kbr}) and~(\ref{muonamm}):
\begin{eqnarray}
|g_{Pq}^{}|^{2}\,\Gamma(X_{P}\to\mu^{+}\mu^{-}) &<& 6.4\times 10^{-29}{\rm ~GeV}
\,\,, \nonumber \\
|g_{Aq}^{}|^{2}\,\Gamma(X_{A}\to\mu^{+}\mu^{-}) &<& 3.0\times 10^{-28}{\rm ~GeV}  \,\,.
\label{combine}
\end{eqnarray}
If  $\Gamma(X_{P,A}\to \mu^{+}\mu^{-})$  are allowed to saturate the bounds of
Eq.~(\ref{muonamm}), the above equations imply that
\begin{eqnarray}
|g_{Pq}^{}|^{2}  \,\,<\,\, 1.7\times 10^{-19} &{\rm ~and~therefore~~}&
{\cal B}(\Sigma^{+}\to p \mu^{+}\mu^{-})
\,\,<\,\, 6.3\times10^{-8}\, {\cal B}(X_{P}\to\mu^{+}\mu^{-})  \,\,,  \nonumber \\
|g_{Aq}^{}|^{2}  \,\,<\,\, 5.8\times 10^{-17} &{\rm ~and~therefore~~}&
{\cal B}(\Sigma^{+}\to p \mu^{+}\mu^{-})
\,\,<\,\, 4.0\times10^{-5}\, {\cal B}(X_{A}\to \mu^{+}\mu^{-})   \,\,.
\hspace*{2em}
\end{eqnarray}
This in turn means that both the pseudoscalar and axial-vector
particles remain viable candidates to explain the HyperCP data after
combining the existing bounds from  $\Delta M_{K_{L}-K_{S}}$,
$a_\mu$, and  \,$K_L\to \mu^+ \mu^-$.\,
In the case of the pseudo-scalar, these combined bounds require that
it decay almost exclusively into a $\mu^+ \mu^-$ pair.

\section{Predictions\label{pred}}

We now turn the argument around and assume that the HyperCP data is indeed explained
by the hypothetical new pseudoscalar or axial-vector particle. This implies that
\begin{eqnarray}
|g_{Pq}^{}|^{2}\, {\cal B}(X_{P}\to \mu^{+}\mu^{-})&=&
\bigl(8.4^{+6.5}_{-5.1}\pm 4.1 \bigr)\times 10^{-20}   \,\,,
\nonumber \\
|g_{Aq}^{}|^{2}\, {\cal B}(X_{A}\to \mu^{+}\mu^{-})&=&
\bigl(4.4^{+3.4}_{-2.7}\pm 2.1 \bigr)\times 10^{-20}  \,\,.
\label{forpreds}
\end{eqnarray}
These can then be used to predict the contributions of the new
particles to other decay modes such as
\,$K \to \pi\pi X_{P,A}\to\pi\pi\mu^+\mu^-$\, and
\,$\Omega^-\to \Xi^- X_{P,A} \to \Xi^-\mu^+\mu^-$.

We first consider  \,$K \to \pi \pi X_{P,A} \to \pi\pi\mu^+\mu^-$.\,
Employing the Lagrangians in Eqs.~(\ref{eff}) and~(\ref{Ls}),
we derive\footnote{We note that each of the $\,\bar K^0\to\pi\pi X_P$\,
amplitudes receives contributions from
both contact and kaon-pole diagrams. The pole terms seem to be
missing in Ref.~\cite{Gorbunov:2000cz}.}
\begin{eqnarray}
{\cal M}(\bar K^0 \to \pi^+\pi^- X_P) &=& {B_0 g_{Pq}^{}\over \sqrt{2}\, f}\,
{\left(m^2_K + m^2_\pi - m^2_{X\pi^-} -  m^2_{\pi^+\pi^-}\right)\over m^2_K-m^2_P}
\,\,,\nonumber\\
{\cal M}(\bar K^0 \to \pi^0\pi^0 X_P) &=& {B_0 g_{Pq}^{}\over 2\sqrt{2}\, f}\,
{\left(m^2_K - m^2_X - m^2_{\pi^0\pi^0}\right)\over m^2_K-m^2_P}
\,\,,\nonumber\\
{\cal M}(K^+ \to \pi^+\pi^0 X_P) &=& {B_0 g_{Pq}^{}\over 2 f}\,
{\left(m^2_{X\pi^-} - m^2_{X\pi^0}\right)\over m^2_K-m^2_P}   \,\,,
\end{eqnarray}
\begin{eqnarray}
{\cal M}(\bar K^0 \to \pi^+\pi^- X_A) &=& -i {\sqrt{2} g_{Aq}^{}\over f}\,
p_{\pi^+}\cdot \epsilon^*  \,\,,\nonumber\\
{\cal M}(\bar K^0 \to \pi^0\pi^0 X_A) &=&  -i { g_{Aq}^{}\over
\sqrt{2}f}\, p_{K}\cdot \epsilon^*  \,\,,\nonumber\\
{\cal M}( K^+ \to \pi^+\pi^0 X_A) &=&  -i { g_{Aq}^{}\over f}
(p_{\pi^0}-p_{\pi^+}) \cdot \epsilon^*   \,\,,
\end{eqnarray}
where  \,$m^2_{ij} = (p_i+p_j)^2$.

Adding the errors in Eq.~(\ref{forpreds}) in quadrature, we obtain the predictions
\begin{eqnarray}
{\cal B}(K_L\to \pi^+\pi^- X_P \to \pi^+\pi^- \mu^+\mu^-) &=&
\bigl(1.8^{+1.6}_{-1.4}\bigr) \times 10^{-9}  \,\,,
\nonumber \\
{\cal B}(K_L\to \pi^0\pi^0 X_P \to \pi^0\pi^0 \mu^+\mu^-) &=&
\bigl(8.3^{+7.5}_{-6.6}\bigr) \times 10^{-9}  \,\,,
\end{eqnarray}
\begin{eqnarray}
{\cal B}(K_L\to \pi^+\pi^- X_A \to \pi^+\pi^- \mu^+\mu^-) &=&
\bigl(7.3^{+6.6}_{-5.7}\bigr) \times 10^{-12}  \,\,,  \nonumber\\
{\cal B}(K_L\to \pi^0\pi^0 X_A \to \pi^0\pi^0 \mu^+\mu^-) &=&
\bigl(1.0^{+0.9}_{-0.8}\bigr) \times 10^{-10}  \,\,.
\end{eqnarray}
Notice that these decay modes are highly suppressed by phase space,
but that at the $10^{-8}$-$10^{-9}$ level they are comparable to existing
limits on other rare $K_L$ decay modes.
The rates for the $K^+$ decay modes are quite sensitive to isospin-breaking
effects, but we find them to be at most at the $10^{-12}$ level, much less
promising than the $K_L$ modes.

We now consider the modes  \,$\Omega^-\to\Xi^-X_{P,A}\to\Xi^-\mu^+\mu^-$.\,
For the pseudoscalar particle, a kaon-pole diagram with vertices from
Eqs.~(\ref{Ls}) and~(\ref{L_PH}) leads to
\begin{eqnarray}   \label{M(O->XP)}
{\cal M}\bigl(\Omega^-\to\Xi^- X_P\bigr)  \,\,=\,\,
\frac{-i B_0^{}\, {\cal C}}{m_K^2-m_P^2}\, g_{Pq}^{}\,
\bar{u}_\Xi^{}q_\mu^{}u_\Omega^\mu   \,\,,
\end{eqnarray}
and for the axial-vector particle a direct vertex from  Eq.~(\ref{L_AH}) gives
\begin{eqnarray}   \label{M(O->XA)}
{\cal M}\bigl(\Omega^-\to\Xi^- X_A\bigr)  \,\,=\,\,
-i {\cal C}\, g_{Aq}^{}\, \bar{u}_\Xi^{}u_\Omega^\mu\, \epsilon_\mu^*  \,\,.
\end{eqnarray}

The resulting branching ratios for
$(\Omega^-\to\Xi^-X\to\Xi^-\mu^+\mu^-)$  are
\begin{eqnarray}  \label{B(O->XP)}
{\cal B}(\Omega^-\to\Xi^- X_P\to\Xi^-\mu^+\mu^-)  &=&
\frac{1}{\Gamma_{\Omega^-}^{}}\,
\frac{\bigl|\bm{p}_\Xi^{}\bigr|^3}{12\pi\, m_\Omega^{}}\;
\frac{B_0^2\, {\cal C}^2\, |g_{Pq}^{}|^2}{\bigl(m_K^2-m_P^2\bigr)^2}
\bigl( E_\Xi^{}+m_\Xi^{} \bigr)\, {\cal B}(X_P\to\mu^+\mu^-)
\nonumber \\
&=&  2.4\times10^{13}\, |g_{Pq}^{}|^2\, {\cal B}(P\to\mu^+\mu^-) \,\,,
\end{eqnarray}
\begin{eqnarray}  \label{B(O->XA)}
{\cal B}(\Omega^-\to\Xi^-X_A\to\Xi^-\mu^+\mu^-)  &=&
\frac{1}{\Gamma_{\Omega^-}^{}}\,
\frac{\bigl|\bm{p}_\Xi^{}\bigr|}{12\pi\, m_\Omega^{}}\; {\cal C}^2\,
|g_{Aq}^{}|^2\,(E_{\Xi}+m_\Xi)\left(3+\frac{\bm{p}_\Xi^2}{m^2_A}\right)
{\cal B}(X_A\to\mu^+\mu^-)
\nonumber \\
&=& 1.6\times 10^{13}\, |g_{Aq}^{}|^2\,{\cal B}(X_A \to \mu^+\mu^-)   \,\,.
\end{eqnarray}
Consequently, the HyperCP data implies
\begin{eqnarray}  \label{B(O->XP->Xmumu)}
{\cal B}(\Omega^-\to\Xi^- X_P\to\Xi^-\mu^+\mu^-)  &=&
\bigl(2.0_{-1.2}^{+1.6}\pm1.0\bigr)\times10^{-6}
\,\,,\nonumber\\
{\cal B}(\Omega^-\to\Xi^- X_A \to\Xi^-\mu^+\mu^-)
&=& \bigl(0.73_{-0.45}^{+0.56}\pm 0.35\bigr)\times10^{-6}  \,\,.
\end{eqnarray}
These numbers represent a substantial enhancement over the existing
standard-model prediction
\,${\cal B}(\Omega^-\to\Xi^-\mu^+\mu^-)=6.6\times10^{-8}$~\cite{Safadi:1987qq}.

\section{Summary}

We have studied the hypothesis that a new particle of mass
$214.3\pm 0.5$\,MeV is responsible for the invariant-mass $m_{\mu^+\mu^-}$
distribution observed by HyperCP in   \,$\Sigma^{+}\to p \mu^{+}\mu^{-}$.\,
We find that existing data on  \,$K^+\to\pi^+\mu^+\mu^-$\,
rule out a scalar particle and a vector particle as possible explanations.
We explore all the existing constraints on pseudoscalar and axial-vector
particles, and conclude that these possibilities are still allowed.
If either one of them is indeed responsible for the HyperCP data,
we predict enhanced rates for  \,$K_L\to \pi\pi X\to\pi\pi\mu^+\mu^-$\,
and  \,$\Omega^-\to\Xi^- X\to\Xi^-\mu^+\mu^-$.\,

\bigskip

{\it Note added}\,\,\, After the completion of our paper, the work
of Deshpande, Eilam and Jiang~\cite{desh} appeared. They reach
similar conclusions to ours.

\begin{acknowledgments}

We thank  HyangKyu Park and Michael Longo for conversations. X.G.H
thanks N.G. Deshpande and J. Jiang for corespondance on their
analysis. The work of X.G.H. was supported in part by the National
Science Council under NSC grants. The work of G.V. was supported
in part by DOE under contract number DE-FG02-01ER41155.

\end{acknowledgments}

\appendix

\section{Derivation of effective Lagrangians\label{Leff}}

The chiral Lagrangian that describes the interactions of the
lowest-lying mesons and baryons is written down in terms of the
lightest meson-octet, baryon-octet, and baryon-decuplet
fields~\cite{GL,Bijnens:1985kj,Jenkins:1991ne}. The meson and baryon
octets are collected into  $3\times3$  matrices $\varphi$ and $B$,
respectively, and the decuplet fields are represented by the
Rarita-Schwinger tensor  $T_{abc}^\mu$, which is completely
symmetric in its SU(3) indices ($a,b,c$). The octet mesons enter
through the exponential $\,\Sigma=\xi^2=\exp( i \varphi/f),\,$ where
$\,f=f_\pi^{}=92.4\rm\,MeV\,$ is the pion-decay constant.

We write the strong chiral Lagrangian at leading order in the
derivative and $m_s^{}$ expansions as
\begin{eqnarray}   \label{Ls}
{\cal L}_{\rm s}^{}  &=& \left\langle \bar{B}^{}\, { i }\gamma^\mu \bigl(
\partial_\mu^{}B+\bigl[{\cal V}_\mu^{},B \bigr] \bigr) \right\rangle
- m_0^{} \left\langle \bar{B}^{} B^{} \right\rangle + D
\left\langle \bar{B}^{} \gamma^\mu\gamma_5^{}
 \left\{ {\cal A}_\mu^{}, B^{} \right\} \right\rangle
+ F \left\langle \bar{B}^{} \gamma^\mu\gamma_5^{}
 \left[ {\cal A}_\mu^{}, B^{} \right] \right\rangle
\nonumber \\ && -\,\, \bar{T}^\mu\,  i \!\not{\!\!{\cal D}}
T_{\mu}^{} + m_T^{}\; \bar{T}^\mu T_{\mu}^{} + {\cal C} \left(
\bar{T}^\mu {\cal A}_\mu^{} B^{}
                 + \bar{B}^{} {\cal A}_\mu^{} T^\mu \right)
+ {\cal H}\; \bar{T}^\mu \!\not{\!\!\!\cal A}\gamma_5^{}\,
T_{\mu}^{} \nonumber \\ && +\,\, \frac{b_D^{}}{2 B_0^{}}
\left\langle \bar{B}^{} \left\{ \chi_+^{}, B^{} \right\}
\right\rangle + \frac{b_F^{}}{2 B_0^{}} \left\langle \bar{B}^{}
 \left[ \chi_+^{}, B^{} \right] \right\rangle
+ \frac{b_0^{}}{2 B_0^{}} \left\langle \chi_+^{} \right\rangle
 \left\langle \bar{B}^{} B^{} \right\rangle
\nonumber \\ && +\,\, \frac{c}{2 B_0^{}}\, \bar T^\mu \chi_+^{}
T_{\mu}^{} - \frac{c_0^{}}{2 B_0^{}} \left\langle \chi_+^{}
\right\rangle
 \bar T^\mu T_{\mu}^{}
\,\,+\,\, \mbox{$\frac{1}{4}$} f^2 \left\langle
D^\mu\Sigma^\dagger\, D_\mu^{}\Sigma \right\rangle +
\mbox{$\frac{1}{4}$} f^2 \left\langle \chi_+^{} \right\rangle
\,\,,
\end{eqnarray}
where  $\,\langle\cdots\rangle\equiv{\rm Tr}(\cdots)\,$  in
flavor-SU(3) space, $m_0^{}$ and $m_T^{}$  are the octet-baryon
and decuplet-baryon masses in the chiral limit, respectively,
$\,{\cal V}^\mu =
\frac{1}{2}\bigl(\xi\,\partial^\mu\xi^\dagger+\xi^\dagger\,\partial^\mu\xi\bigr)
+ \frac{i}{2}\bigl(\xi^\dagger\ell^\mu\xi+\xi
r^\mu\xi^\dagger\bigr),\,$ $\,{\cal A}^\mu  =
\frac{ i }{2}\bigl(\xi\,\partial^\mu\xi^\dagger-\xi^\dagger\,\partial^\mu\xi\bigr)
+ \frac{1}{2}\bigl(\xi^\dagger\ell^\mu\xi-\xi
r^\mu\xi^\dagger\bigr),\,$ $\,{\cal D}^\mu T_{klm}^\nu =
\partial^\mu T_{klm}^\nu + {\cal V}_{kn}^\mu T_{lmn}^\nu
+ {\cal V}_{ln}^\mu T_{kmn}^\nu + {\cal V}_{mn}^\mu
T_{kln}^\nu,\,$ $\,D^\mu\Sigma =\partial^\mu\Sigma + { i }\ell^\mu\Sigma
- i \Sigma r^\mu,\,$ $\,\chi_+^{} =
\xi^\dagger\chi\xi^\dagger + \xi\chi^\dagger\xi,\,$ with
$\,\ell^\mu=\frac{1}{2}\lambda_a^{} \ell_a^\mu = v^\mu + a_\mu$,\,
$\,r^\mu=\frac{1}{2}\lambda_a^{}r_a^\mu = v^\mu - a^\mu$,\,  and  \,$\chi=s+i p$\,
containing external sources.  In the absence of external sources,  $\chi$
reduces to the mass matrix $\,\chi=2B_0^{}\, {\rm diag}(\hat{m},\hat{m},m_s^{})=
{\rm diag}\bigl(m_\pi^2,m_\pi^2,2m_K^2-m_\pi^2\bigr)\,$ in the
isospin-symmetric limit  $\,m_u^{}=m_d^{}=\hat{m}$.\,\, The
constants  $D$, $F$, $\cal C$, $\cal H$, $B_0^{}$, $b_{D,F,0}^{}$, $c$, and
$c_0^{}$ are free parameters which can be fixed from data.

To extract the couplings of the new particles $X$ from the above Lagrangian,
we identify the external sources with
\begin{eqnarray}
s &=& g_{Sq}^{} X_S^{}\, (T_6 + i T_7)  \,+\,{\rm H.c.} \,\,, \nonumber \\
p &=&   g_{Pq}^{} X_P^{}\, (T_6 + i T_7) \,+\,{\rm H.c.}\,\,,\nonumber \\
v^\mu &=&  g_{Vq}^{} X_V^\mu\, (T_6 + i T_7)  \,+\,{\rm H.c.}\,\,,\nonumber \\
a^\mu &=&   g_{Aq}^{} X_A^\mu\, (T_6 + i T_7) \,+\,{\rm H.c.}  \,\,.
\end{eqnarray}
The Lagrangians in Eq.~(\ref{eff})  then follow.

Numerically, we adopt the tree-level values  $\,D=0.80\,$  and
$\,F=0.46,\,$ extracted from hyperon semileptonic decays, as well
as  $\,|{\cal C}|=1.7,\,$ from the strong decays  $\,T\to
B\varphi.\,$ Furthermore, using  $\,\hat{m}+m_s^{}=121\,{\rm MeV}$\, and
isospin-symmetric values of the
baryon and meson masses, we have
\begin{eqnarray}
b_D^{}  \;=\;  0.270   \,\,,
\hspace{3em}
b_F^{}  \;=\;  -0.849   \,\,,
\hspace{3em}
B_0^{}   \;=\;  2031\;{\rm  MeV}   \,\,,
\end{eqnarray}
the other parameters being irrelevant to our calculations.

\end{document}